# INDUCED CRYSTALLIZATION METHOD OF RAPID METAL MELTS 3D-PRINTING

V.B.Oshurko, V.N.Lednev, A.M.Mandel, K.Solomakho

*Prokhorov General Physics Institute of RAS,*

*Vavilov st. 38, Moscow, Russia*

*Moscow State Technological University "Stankin", 127994, Vadkovsky 1,*

*Moscow, Russia*

***Abstract.*** *A solution for the most important problems in 3D printing technology (long print times, small build volumes and limited material properties) is proposed. Method is based on the fact that temperature of melts crystallization can be shifted up by pressure. Required pressure can be created by Ampere force caused by electric current in the melt flow. Magnetohydrodynamical model (with heat taken into account) is developed. It has been shown that additional heating of melt by this current can be overcome by appropriate choice of tunable parameters and introduction of some additional magnetic field. Range of parameters where this induced crystallization can take place is found.*

**Intorduction**

It is well known that three-dimensional (3D) printers today are rarely used for industrial manufacturing. Currently, the greatest problems in 3D printing technology are long print times, very small build volumes, and limited material properties. For example, typical productivity of laser sintering (powder materials) is about tens of grams per hour or tens of grams per minute for cold spray. Another serious problem is rather bad quality of sintered polycrystalline material.

There are many works ([1],[2]) directed towards solution of these problems. In [1] a very promising method of RLP (rapid liquid printing) is proposed. This method is based on liquid material deposition into the granular gel to form 3D structures. However, this method is hardly applicable to solution of the most interesting industrial problem of liquid metal 3D-printing (e.g. steel melt, etc).

In present work a new simple principle of field-induced crystallization (FIC) for rapid liquid printing for metal melts is proposed. The idea of our method is as follows. Consider continuous flow of metal melt from tank or reservoir through appropiate nozzle. When electric current is applied to this continuous flow (continuous jet) of melted metal, the flow should be compressed by well-known Ampere force. Obviously, pressure inside the flow is controlled by the electric current. On the other hand, the increasing of pressure leads to the increase of crystallization temperature [3]. If the pressure in the region of the end of the flow is high enough, then the transition through the crystallization point to a solid state would occur in this region. If this crystallization takes place, bulding of appropriate 2D-scanning procedure and excess material removal procedure leads to creation of rapid 3D printer for metal melts.

However, there is a process that obviously can prevent cristallization in described process. Indeed, the increase of electric current (which is necessary to produce high pressure) leads to strong additional heating of the flow. On the other hand, there are two processes that can cool the flow and hence work 'against' heating by the electirc current. First, in the case of metal melts, the most effective cooling process is radiational one. Indeed, energy loss by radiation is proportinal to $\sim:T^4$ (where $T$ is absolute temperature) and in the case of melts ($T \approx 10^3 K$) could be rather efficient. This process depends on radiating area, e.g. on geometry of flow. Another cooling process is the income of 'unheated' melt from the resevoir due to motion of melt in the flow. Efficiency of cooling in this process depends on flow velocity.

Thus, it is not clear *a priori* is there a region of tunable parameters (flow velocity, electric current, melt temperature, nozzle diameter) where the described crystallization can take place. The purpose of present work is to build a theoretical model of all these processes to analyze the possibility of controlled crystallization in described conditions. It will be shown below that after some modification of procedure (introduction of additional magnetic field) this method requires realistic values of tunable parameters.

**Methods: magnetohydrodynamical model**

Consider liquid metal flow (jet) along the $z$ axis in downward direction at initial velocity of $v_0$. Let the flow be a stationary process, so that $d\vec{v}/dt = 0$. It means that usual (stationary) magnetohydrodynamics equation, which is Navier-Stokes equation with magnetic term can be used [4]:

$$(\vec{v}\nabla)\vec{v} = \nu\Delta\vec{v} - \frac{1}{\rho}\nabla p + \vec{g} + \frac{1}{\rho}[\vec{j},\vec{B}] \qquad (1)$$

where $\vec{v}$ is the speed, $p$ - pressure, $\nu$ - viscosity, $g$ - acceleration of gravity, $\vec{j}$ - current density and $\vec{B}$ - a magnetic field.

In our case magnetic field is not external and is caused by electric current in the jet. The magnitude of this magnetic field could be easily estimated. Let electric current density be constant over the cross section of the jet. Strictly speaking, electric current density distribution over cross section of the jet should be found from energy minimization condition (see for example, [4]). However, this could change our estimation only by a constant coefficient ~ 1. Since $\vec{B} = \mu_0\mu[\vec{j},\vec{r}]$, we obtain for magnetic field radial distribution ($r < R$, where $R$ is jet radius)

$$B(r) = \mu\mu_0 jr/2, \tag{2}$$

where the current density

$$j = I_0/\pi R^2(z) \tag{3}$$

Here $\mu$ is the magnetic permeability of a liquid metal, $\mu_0$ is the magnetic constant, $I_0$ is the current value. In cylindrical coordinates the last term in equation (1) has only a radial component equal to

$$-\frac{\mu\mu_0}{\rho} rj^2 \tag{4}$$

which is the Ampere force (divided by density $\rho$). As expected, the Ampere force acts along the radius and directed to axis of the jet ($r = 0$), i.e. in direction of compression. We assume that the jet regime is very close to laminar one and the diameter of the jet is smaller than its length. Also we neglect the angular and radial components of the total flow velocity $\vec{v} = (v_r, v_\phi, v_z)$, so that $v_r \approx 0, v_\phi \approx 0$ and further we denote $v_z = v$.

We do not use traditional assumption of the incompressibility of liquid, because rather high pressures could be achieved in our case (100-1000 kbar) if the electric current is high enough. In addition, the assumption of the incompressibility of a fluid $div(\vec{v}) = 0$ would lead to $dv_z/dz = 0$ or to $v_z = Const$. Hence, if the continuity equation $Q_0 = \rho v(z) S(z)$ is taken into account (where $Q_0$ is the mass flux, $kg/s$, $S(z)$ is the jet cross-section) then the incompressibility leads to constant jet cross-section $S(z)$ along $z$.

As a result, in cylindrical $(r,\theta,z)$ coordinates, the equation for pressure dependence on $r$ is obtained (for $\vec{e}_r$)

$$\frac{1}{\rho}\frac{\partial p}{\partial r} = -\frac{\mu\mu_0}{\rho} rj^2 \tag{5}$$

Equation for $\vec{e}_z$ gives $v$

$$v\frac{\partial v}{\partial z} = \nu\frac{\partial^2 v}{\partial z^2} - \frac{1}{\rho}\frac{\partial p}{\partial z} + g \tag{6}$$

The next approximation is that $v_z = v$ does not depend on radius $r$. Indeed, this is obviously good approximation for the free flow in the absence of fixed walls or tube (in case of tube the flow velocity would have to be zero at the wall). Then, according to equation (6), the pressure $p = p(z)$ in the right-hand side of equation (6) also should not depend on the radius, which contradicts equation (5) for $p$. This contradiction disappears if we assume that both $p$ and $v_z$ also weakly depends on $r$. Then we could use $p$ averaged over $r$ at any given coordinate $z$. Integrating equation (5), we obtain for $p(z,r)$:

$$p(z,r) = -\mu_0\mu J^2 \frac{r^2}{2} + C_1 \tag{7}$$

where $C_1$ is the integration constant. Assuming the weak dependence of $p(r)$, we should take $p(z)$ as the average value (on the area element $2\pi r dr$) over the area $\pi R(z)^2 = S(z)$. The condition of continuity of the electric current $J(z)S(z) = I_0$ is also have to be taken into account:

$$p(z) = \frac{1}{\pi R^2}\int_0^R p(z,r)2\pi r dr = \frac{\mu_0 \mu I_0^2}{16\pi S(z)} + C_1 \qquad (8)$$

where the integration constant is just the static pressure.

With regard to continuity of substance flow $\rho v_0 S_0 = \rho v(z)S(z)$ (here $v_0$ and $S_0$ respectively, the initial velocity and the jet section area at $z=0$), the equation (6) can be easily solved. However, exact solution is expressed in terms of Bessel functions or Airy functions. As a result, expressions for constants are transcendental equations and could be solved only numerically. To obtain simpler qualitative relations between parameters we could consider very simplified problem by putting the viscosity in equation (17) equal to zero.

In this case, equation (17) is simply the differential form ($d/dz$) of well-known Bernoulli equation [5]. It is easy to see that the pressure caused by magnetic field of electric current is nothing more than the energy density of the magnetic field, as it must be in Bernoulli equation. Thus, our model has is the same applicability as the Bernoulli equation. Hence, this equation, modified with respect to magnetic field, could be written in the form

$$\rho v^2(z)/2 + \rho g z + P_0 + p(z) = const, \qquad (9)$$

where $p(z)$ is the pressure caused by magnetic field and actually equal to the energy density of magnetic field in the cross section at the coordinate $z$, respectively.

Now we have to define controlled (e.g. tunable in experiment) parameters of this model. Obviously, following parameters could be changed in experiment : the total electric current $I_0$, the initial section of the jet, i.e. cross-section ares of the nozzle $S_0$, the length of the jet $L$ and the pressure $P_0$. It is easy to see that pressure $P_0$ defines the initial velocity $v_0$. Further we can consider the speed $v_0$ as tunable parameter.

First of all, we need to find the pressure $P_1$ at the end of the jet as the function of these tunable parameters. Substitute the cross-section at the end of the jet $S_1$ (i.e. at a given distance $L$ into equation (8). For this purpose the Bernoulli equation (9) can be written as

$$\frac{1}{2}\rho v_0^2 + \rho g L + \frac{\mu_0 \mu I_0^2}{16\pi S_0} = \frac{\rho v_0^2 S_0^2}{2 S_1^2} + \frac{\mu_0 \mu I_0^2}{16\pi S_1} \qquad (10)$$

where the left-hand side refers to the upper point of the jet, and the right-hand side to the lower point; we've replaced $v(L) = v_0 S_0/S_1$ due to the continuity condition. Now it is easy to find section area $S_1$ and corresponding pressure $P_1$:

$$S_1 = \frac{S_0\left(\dfrac{1}{16\pi}\mu_0\mu I_0^2 + \sqrt{A}\right)}{\left(\rho v_0^2 S_0 + 2\rho g L S_0 + \dfrac{1}{8\pi}\mu_0\mu I_0^2\right)} \quad (11)$$

$$P_1 = \frac{\mu_0\mu I_0^2\left(\rho v_0^2 S_0 + 2\rho g L S_0 + \dfrac{1}{8\pi}\mu_0\mu I_0^2\right)}{16\pi S_0\left(\dfrac{1}{16\pi}\mu_0\mu I_0^2 + \sqrt{A}\right)} \quad (12)$$

где

$$A = (\rho^2 v_0^2 + 2\rho^2 gL)S_0^2 v_0^2 + \left(\frac{1}{8\pi}\rho v_0^2 S_0 + \frac{1}{256\pi^2}\mu 0\mu I_0^2\right)\mu_0\mu I_0^2 \quad (13)$$

**Heat balance**

Since crystallization is the most important process considered here, the distribution of temperature in the melt flow is of fundamental importance. Let's consider the temperature regime of the jet of liquid metal under current. It is clear that this regime is determined by the balance of four heat fluxes. First of all, the metal melt can cool (lose the energy) due to radiation to the environment. In case of sufficiently high temperatures (about or higher than the temperature of melting), such losses might not be smal . We could write for this loss of energy (heat transfer) of the jet element with height $dz$ and radius $R(z)$ per unit time (here we neglect the temperature of medium comparing to the melt temperature)

$$dQ_1 = \sigma T^4 \cdot 2\pi R dz, \tag{14}$$

where $\sigma$ is the Stefan-Boltzmann constant. This heat dissipation is compensated by the heat release due to electric current:

$$dQ_2 = I_0^2 dz / \chi S(z). \tag{15}$$

There are also two processes that can "equalize" the temperature distribution along the jet. The first one, the "convection-like" heat flux due to melt income from the reservoir:

$$dQ_3 = \rho C_p \cdot v(z) S(z) dT = \rho C_p \cdot v_0 S_0 dT, \tag{16}$$

we have used the continuity equation of matter; $C_p$ is the specific heat of metal here. The second one, is usual heat diffusion due to thermal conductivity:

$$dQ_4 = \kappa \frac{S(z)}{dz} \frac{\Delta T}{dz/L} \tag{17}$$

As a result Fourier equation for the heat balance can be written as:

$$dQ_4 + dQ_3 + dQ_2 - dQ_1 = 0 \tag{18}$$

Let's estimate contributions of all the components of this equation. Following parameters of steel (for example) have been used in present work: $T_m$ = 1790 K, density $\rho$ = 7000 kg/$m^3$, thermal conductivity $\kappa$ = 47 W/m/K, specific heat Cp = 825 J/kg/K, electrical conductivity $\chi = 10^7 Ohm^{-1} m^{-1}$ [6]. Also, Stefan-Boltzmann constant is $\sigma = 5.67 10^{-8} W/m^2/K^4$ and the magnetic permeability of the melt must be about $\mu \approx 1$ because the melt temperature is always higher than Curie temperature. Other parameters used in estimations are : $I_0$ =1..1000 A, $v_0$ =0.1-100 m/s, $\Delta T$ =1..100 K, $dz$ =0.001 m, $L$ =0.01..1 m. It has been estimated that the heat flux due to the thermal conductivity $Q_4$ is much smaller (several orders of magnitude) than the rest components ($Q_1, Q_2, Q_3$) in wide range of experimental conditions. Thus, it is possible to neglect the contribution of thermal conductivity with respect to heating by electric current and radiation losses. Then the heat balance equation (divided by $dz$) takes the form:

$$\frac{\Delta T}{\Delta z} = \frac{1}{\rho C_p v_0 S_0}\left[\frac{I_0^2}{\chi S(z)} - 2\pi\sigma R(z)T^4(z)\right]. \qquad (19)$$

The equation has no analytical solution, but it could be numerically integrated by substitution of explicit functions $S(z)$ and $R(z)$ from the preceding equations. However, our purpose is to analyze the possibility of controlled crystallization in the melt flow and it is not necessary to search for solution of this equation.

It should be mentioned, that the heat flux $Q_3$ associated with the melt income performs a stabilizing function. If the temperature of the jet is lower than the temperature in melt reservoir, then the heat flux is positive and heating occurs. In the opposite case, this stream could cool the jet on the contrary. Since the magnitude of this flux is determined by the initial velocity of liquid it is possible to control this heat flux by changing this initial velocity $v_0$. In other words, increasing the velocity of the jet can compensate excessive heating.

### Results and discussion

The main purpose of our study is to find out is there a range of tunable (control) parameters ($v_0$, $I_0$, $S_0$, $L$) where current-controlled melt crystallization at the end of jet could occur. For this purpose, we have to estimate pressure value which is necessary for shifting the crystallization temperature up. It is known ([3],[7]) that this shift of crystallization temperature can take place at rather high pressures, on the order of hundreds of kbar. For steel, empirical dependence of melting point on pressure could be expressed as ([7])

$$T_c r = T_m + \alpha \cdot p \qquad (20)$$

where $T_m$ is the melting point under normal conditions; for steel it could be taken [6] $T_m$ = 1796 K, $\alpha = 2.4 \cdot 10^{-8}$ Pa/K. It means that the pressure of about 100 kbar is required to change the melting point by 1 K.

To be definite, we would assume that we need to increase the crystallization temperature by $\delta T$ = 1 K. This increase can be achieved, according to (20), at some pressure $P_{min} = \delta T/\alpha$. This gives a condition on the pressure $P_1 > P_{min}$ and it could be transformed to the condition for the current $I_0 > I_{min}$. In other words, the crystallization takes place if the current is greater

than some $I_{min}$. Then $I_{min}$ is the solution of equation

$$\delta T/\alpha = P_1(v_0, I_0 = I_{min}, S_0, L) \qquad (21)$$

where $P_1$ is given by the expression (12). The solution of this algebraic equation of fourth degree in $I_0^2$ is trivial, but is too large to be put here. The result as a function of $I_{min}(v_0)$ is shown in Fig.1 (a) - Fig.1 (c).

Another condition follows from the balance of heat fluxes. If electric current producing this pressure creates the heat flux $I_0^2 dz/\chi S_1$ greater than the heat loss $\sigma T^4 2\pi (S_1/\pi)^{1/2} dz$, that does not mean obligatory the increase of temperature. Indeed, according to heat balance equation, in this case "convective" term (due to income flow) would lead to decrease of heating. Then, in order for the heating to be still smaller than the increase in the temperature of crystallization $\delta T$, the necessary requirement is that the initial velocity of the flow in the "convective" term must be greater than some minimum velocity $v_0 > v_{0min}$. This velocity $v_{0min}$ would be determined by the condition $\Delta T/\Delta z < \delta T/L$, from which the expression for calculating $v_0$ can be obtained:

$$\frac{\delta T}{L} = \frac{1}{\rho C_p v_0 S_1}\left(\frac{I_0^2}{\chi S_1} - 2\sigma T_m^4 \pi\sqrt{\frac{S_1}{\pi}}\right) \qquad (22)$$

If we assume that velocity $v_0$ is given, then the condition $v_0 > v_{0min}$ could be expressed as condition on maximum current ($I_0 < I_{max}$) at which cooling by melt income is still enough to provide $\Delta T/\Delta z < \delta T/L$. Then expression (22) defines the function $I_{max}(v_0, S_0, L)$. The result is elementary got, but it's also too large to be presented here entirely. Calculation of the current $I_{max}$ as a function of the velocity $v_0$ is also shown in Fig.1 (a) - Fig.1 (c).

Now we have two functional dependencies to find out if the described crystallization mechanism exists: 1, $I_{min}$ - i.e. the minimum electric current at which the pressure is already sufficient to increase the crystallization temperature by $\delta T$ and, 2, the current $I_{max}$ is the maximum current at which heating by this current is still compensated by the oncoming flow of the colder melt. The ranges of the parameters for which $I_{max} > I_{min}$ are obviously the required conditions for the implementation of the proposed method.

The results of this calculation for the case of steel melt are shown in Fig. 1 (a). As could be seen, there are no conditions where $I_{min} < I_{max}$. In other words, it's impossible to compensate ohmic heating by current (if the current is sufficient for change of crystallization temperature) due to flow of cooler melt.

However it is easy to see that the desired pressure could be reached at much lower current level if the external magnetic field is added. This external field must have the same geometry as the field of electric current to provide compression of the jet. Hence the vector $\vec{B}$ must always be directed tangentially to the circle around the axis of jet. This tangential field could be produced, for example, by a toroidal-shaped solenoid (or a segmented toroidal solenoid, i.e. a toroidal winding with spaces) located around the lower part of the jet. In this case, additional magnetic field is indeed tangential and it is possible to use for the force $\vec{F} = [\vec{j}, \vec{B}]$. According to (1) additional pressure can be written as

$$P_{ext} = \frac{I_0 B}{S_1^2} \qquad (1)$$

where $S_1$ - is the cross section of jet at the bottom of the flow.

The purpose of our study is to clear up the possibility of realizing of described technique. Therefore it is possible to omit the questions of the optimal shape of the toroid section, location, and so on. Assume that described tangential field with induction $B$ is created. Functions $I_{max}(v_0)$ and $I_{min}(v_0)$ at different values of $B$ are presented in Fig. 1 (b) - Fig. 1 (d). as can be seen in this figures, results have changed significantly when a relatively small additional field (of only a few tens of millitesla ($B$ = 50 mT)) is applied. As can be seen from Fig. 1 (b), there is the fairly large area where $I_{max} > I_{min}$ with quite realistic values of tunable parameters: speeds above $3 m/s$ and a electric current greater than $1300 A$. Fig. 1 (c) and Fig. 1 (d) show how this region varies with the change in other control parameters - jet length $L$ and hole diameter $D$. It could be seen from Fig. 1 (c) that the increase in jet length up to 5 cm with approximately the same currents requires the increase in the initial velocity $v_0$ to $9.5 m/s$.

It is easy to see that the diameter of the nozzle hole is the principal parameter for this method. If the diameter increased to $6.5 mm$ (Fig. 1 (d)) then the region where $I_{max} > I_{min}$ does not exist.

The most important fact is that the increase of additional magnetic field can significantly reduce values of minimal current and minimal velocity. Thus, there is wide range of practically real values of parameters at which the described method of 3D printing can be realized.

However, two significant remarks to the proposed method should be made. First, we have neglected the viscosity of the melt in the described model. How much could it change the described picture if viscosity would be taken into account? Non-zero viscosity could decrease the initial flow velocity through the hole (there is so-called "hole ratio", which is determined by viscosity). However, the lack of speed due to viscosity of melt could be easily compensated by increasing the pressure in the original reservoir $P_0$. In addition, viscosity of melts is usually rather small (~ $:10^{-7}$ Pa·s). It can be shown that it leads to insignificant changes in velocity along short path (1.8 cm) of the jet in our case.

The second remark is that the latent heat of crystallization has not been taken into account in our calculations. This additional heat could be released during crystallization and change the picture of the heat balance. However, it is easy to estimate that this heat is much less than ohmic heating at currents of several thousand amperes. In addition the small excess of heat could be also always compensated by the choice of $v_0$ and $I$ in the experiment within the range of permissible parameters.

**Conclusions** It has been shown that desribed method of field-induced crystallization (FIC) at the end of melt jet cannot be realized directly (for steel melts) because the heating of melt (by current) is too large. However, it has been found that addition of external tangential magnetic field can change the situation. In this case, there is wide range of practically real values of parameters where FIC method of 3D printing can be realized.

**Acknowledgements**

The work has been supported by the Ministry of Education and Science of the Russian Federation (grant 9.1195.2017/4.6).

**Figure legends**

Fig.1. Region of electric currents $I$ and flow velocities $v_0$ where $I_{max} > I_{min}$ exists (b),(c) and does not exist (a),(d). $I_{max} > I_{min}$ means that heating by current can be compensated by cooler melt income and, hence, field-induced crystallization is possible. Parameters: (a) jet length is 1.8 cm,

nozzle diameter is 0.35 cm, no external field; (b) jet length is 1.8 cm, nozzle diameter is 0.35 cm, external field 50 mT; (c) jet length is 5.0 cm, nozzle diameter is 0.35 cm, external field 50 mT;(d) jet length is 1.8 cm, nozzle diameter is 0.65 cm, external field 50 mT.

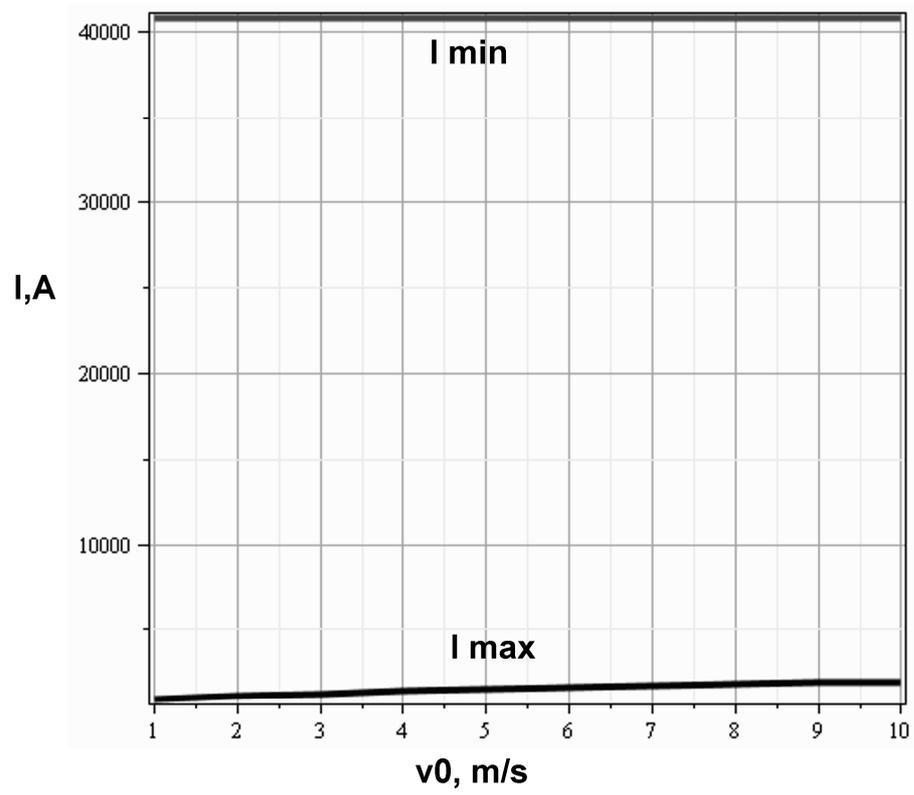

1a

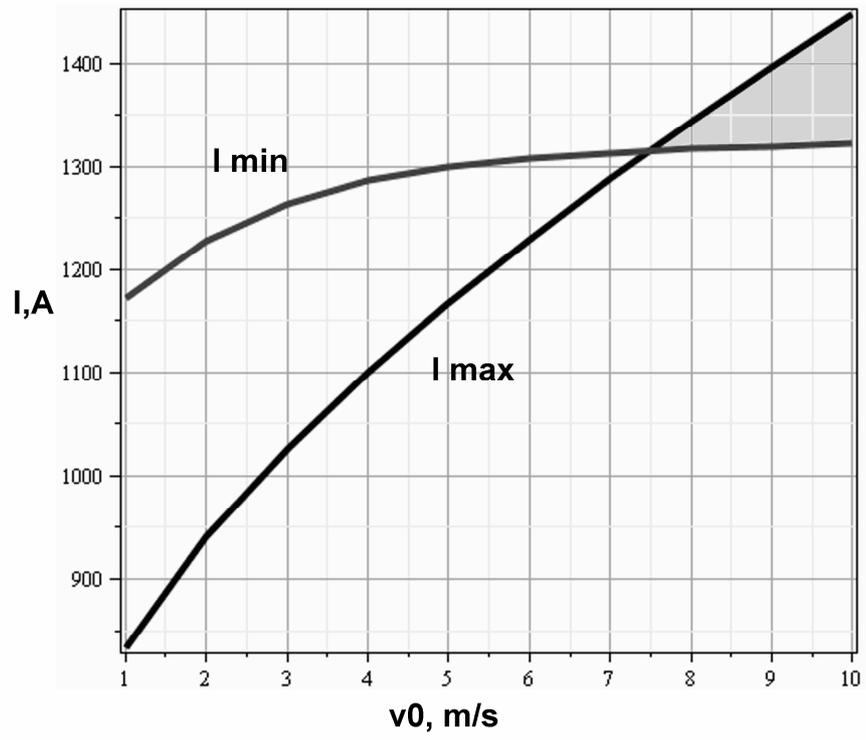

1b

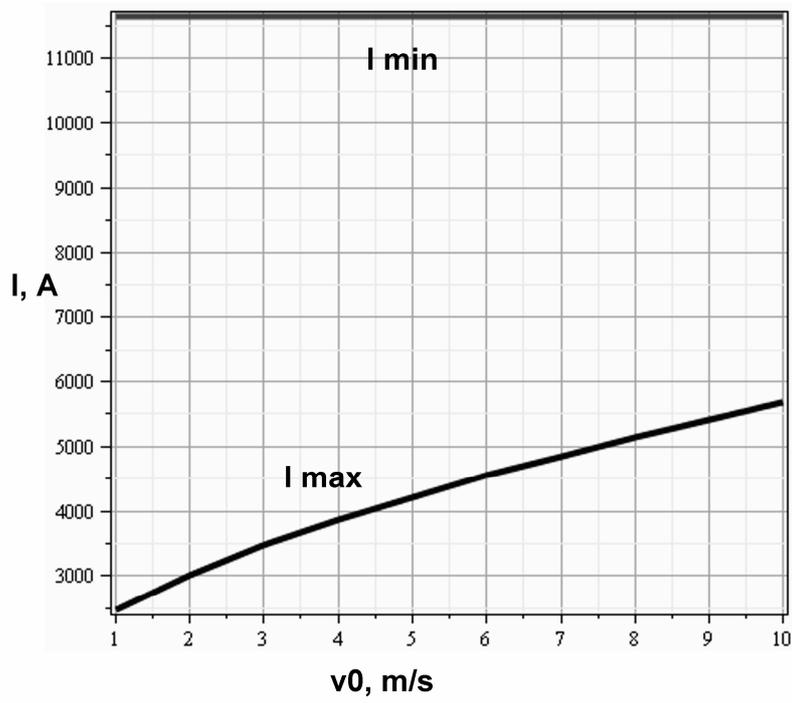

1c

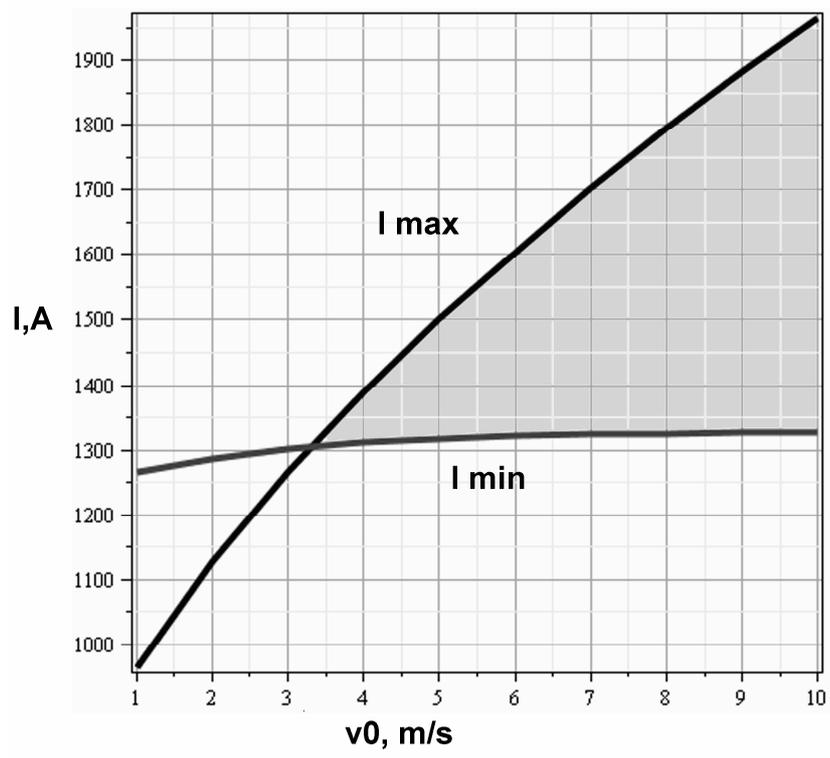

1d